\documentclass[11pt,twoside]{article}

\usepackage{asp2006}
\usepackage{epsf}
\usepackage{psfig}
\usepackage{lscape}
\usepackage{natbib}
\usepackage{graphicx}

\begin{document}
\title{Recurrent Novae: Progenitors of SN Ia?}   

\author{Rolf Walder and Doris Folini} 
\affil{\'{E}cole Normale Sup\'{e}rieure, Lyon, CRAL, UMR CNRS 5574, 
       Universit\'{e} de Lyon, France} 

\author{Jean M. Favre} 
\affil{Swiss National Supercomputing Centre (CSCS), 
       CH-6928 Manno, Switzerland}    

\author{Steven N. Shore} 
\affil{Dipartimento di Fisica "Enrico Fermi", 
       Università di Pisa and INFN-Sezione di Pisa, 
       Pisa 56127, Italy}    

\begin{abstract} 
We present 3D hydrodynamical simulations of the separated binary RS
Ophiuchi (RS Oph), a recurrent nova and potential progenitor of a SN
Ia. RS Oph is composed of a red giant (RG) and a white dwarf (WD)
whose mass is close to the Chandrasekhar limit. In an isothermal
scenario, the WD accrets about 10\% of a 20~km/s RG wind by a
non-Keplerian accretion disk with strong spiral shocks, and about
2\% of a 60~km/s RG wind by what we term a 'turbulent accretion ball'.  A
significantly larger impact have the thermodynamics. In an adiabatic
scenario only about 0.7\% of the 20~km/s RG wind is accreted.  The
rate of change of the system separation due to mass and angular
momentum loss out of the system is negative in all three cases
studied, but is ten times smaller for a fast RG wind (60 km/s) than for
a slow RG wind (20 km/s). The results demonstrate that existing nova
models and observed recurrence times fit well together with 3D wind
accretion and that RS Oph is one of the most promising systems to
become an SN Ia.
\end{abstract}

\section{Introduction}
\label{sec:intro}
Type Ia supernovae (SNe Ia) are cornerstones of modern cosmology as a
measure for cosmological distances, and they are crucial building
blocks of the universe, as production sites of a large part of iron
group elements. In an SN Ia a white dwarf (WD) explodes after his mass
has surpassed the Chandrasekhar stability limit.  The existence of SN
Ia thus is closely linked to the question of mass gain of the WD by
accretion. The layer of accreted matter occasionally undergoes a
thermonuclear runaway (nova) and much of the previously accreted
matter is blown into space again.  Based on 1D
models, \cite{yaron-et-al:05} find that a net mass gain of an already
massive WD ($M \approx 1.4 M_\mathrm{\odot}$) occurs only for accretion
rates above roughly $10^{-8} M_\mathrm{\odot}$yr$^{-1}$.  Moreover, to
achieve a net gain of 0.05 $M_\mathrm{\odot}$ the WD must undergo many
nova cycles, requiring a total time on the order of 1 to 10 million
years.

Particularly interesting binary star systems in this context are
recurrent novae~\citep{anupama:02}. From the short decay time of their
nova light curves the existence of an already massive WD close to $1.4
M_\mathrm{\odot}$ can be inferred. In the subclass of symbiotic-like
recurrent novae, which we are interested in here, the secondary is a
red giant (RG) star which undergoes a substantial mass loss in the
form of a stellar wind. The RG typically does not fill its Roche lobe
and mass transfer onto the WD occurs through wind accretion. The
accretion efficiency then is on the order of 10\% of the RG mass
loss. Nevertheless, with a physically plausible RG mass loss rate of
$10^{-7} M_\mathrm{\odot}$yr$^{-1}$ the necessary WD accretion rates
can be reached. Sustaining these conditions over several million years
is not in a priori contradiction with typical RG life
times. Symbiotic-like recurrent nova thus are plausible candidates for
SN Ia progenitors, but a number of questions remain open.

The best observed system of this kind is RS Ophiuchi (RS Oph), which we
investigated already in ~\citet{walder-folini-shore:08} where we
presented a 3D hydrodynamical simulation of the recurrent nova RS Oph
during the accretion phase and nova outburst. Two issues we take up
here again, in the light of new simulation data: accretion rates in a
recurrent nova binary star system and the competition between mass
loss and angular momentum loss out of the system. The latter is
crucial as it affects the separation of the system and thus the
accretion rate.
\section{Model Problem and Numerical Method}
\label{sec:model}
The symbiotic-like binary star system RS Oph consists of a red giant
and a white dwarf. The orbital period is 455
days~\citep{fekel-et-al:00, dobrzycka-kenyon:94}. The system undergoes
a nova outburst about every 22 years~\citep{anupama:02}. For the most
recent outbursts in 1985 and 2006, exquisite panchromatic
observational data are available
\citep{shore-et-al:96, bode-et-al:06, sokoloski-et-al:06,
das-et-al:06, obrien-et-al:07, 2006MNRAS.373L..75E, worters-et-al:07,
hachisu-et-al:07, bode-et-al:07}. We adopt masses of 2.3
$M_\mathrm{\odot}$ and 1.4 $M_\mathrm{\odot}$ for the RG and WD,
respectively and a corresponding separation between the components of
$a = 2.68\cdot 10^{13}$~cm. We assume phase locking of the RG, a RG
terminal wind velocity of either $v_\mathrm{RG} = 20$~km/s (simulation
slow) or $v_\mathrm{RG} = 60$~km/s (simulation fast) in the rest frame
of the RG, a mass loss rate of $10^{-7}$ $M_\mathrm{\odot}$yr$^{-1}$,
and a wind temperature of 8000~K. The radius of the RG is smaller than
its Roche lobe, accretion by the WD occurs from the RG wind.

We solve the Euler equations in 3D with a nearly
isothermal polytropic equation of state with $\gamma =1.1$, resulting
in a thermal structure that comes close to that obtained from
photoionization models of related symbiotic binary
systems~\citep{1987A&A...182...51N, nussbaumer-walder:93}. For
comparison, we also did one adiabatic simulation with $\gamma
=5/3$ and $v_\mathrm{RG} = 20$~km/s. 

All simulations were performed with the hydrodynamical codes from the
A-MAZE package~\citep{walder-folini:00}. The accreting system was
relaxed to a quasi-stationary state. Mass and angular momentum losses
out of the system were then computed through spherical shells around
the center of mass. For further details on the simulation set up we
refer to~\citet{walder-folini-shore:08}.
\begin{figure}[ht]
\begin{center}
\includegraphics[width=13cm]{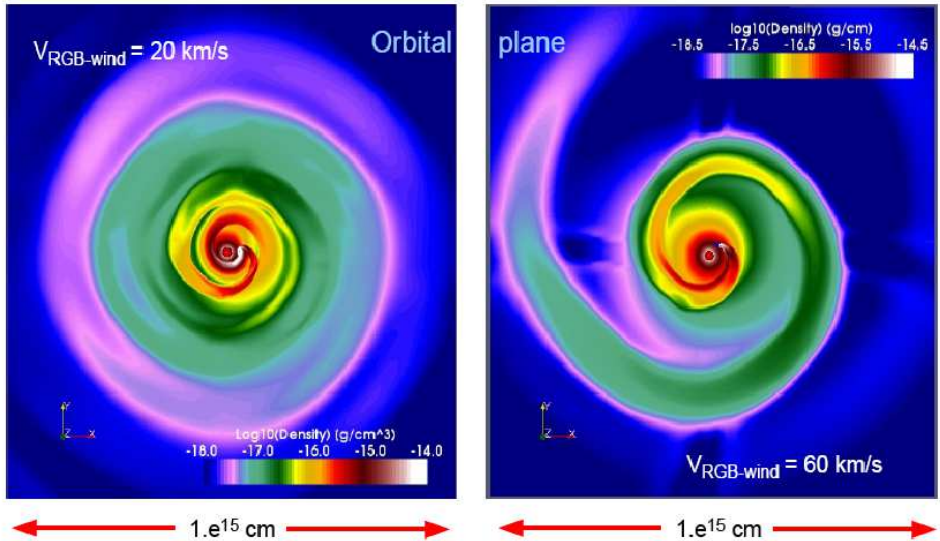}
\end{center}
\vspace{-0.5cm}
\caption{Density in orbital plane for slow (left) and fast (right) simulation.}
\label{fig:spiral}
\end{figure}
\section{Large scale patterns, accretion, system evolution}
\label{sec:characteristics}
During the roughly 22 years of quiescence between successive nova
outbursts, the circumstellar environment gets filled with RG material,
which ultimately reaches out to several $10^{15}$ cm.  The matter
distribution on these large scales differs markedly between the slow
and fast simulation, as shown in Fig.~\ref{fig:spiral}.  In the slow
case, the situation is fairly smooth and spherically symmetric. The
spiral pattern in the equatorial plane is rather faint. A density
contrast of a factor 2-3 between polar (less dense) and equatorial
directions exists. This is sufficient to transform a symmetric nova
outburst of the WD into an asymmetric nova remnant with an aspect
ratio of about 2:1~\citep{walder-folini-shore:08}. The observed nova
remnant in RS Oph is asymmetric as well~\citep{obrien-et-al:07}, a
quantitative comparison remains to be done. For somewhat simpler
underlying models, observable quantities have already been modeled and
successfully compared with
observations~\citep{vaytet-et-al:07,orlando-et-al:09}.  The fast
simulation shows a much more pronounced, farther reaching spiral
structure in the equatorial plane (Archimedian spiral). Density
contrasts across spiral arms exceed one order of magnitude even at
distances ten times the system separation. Densities in the polar
direction are much more smooth.
%
%

\begin{figure}[ht]
\begin{center}
\centerline{\hspace{10.5cm}
\includegraphics[width=26cm]{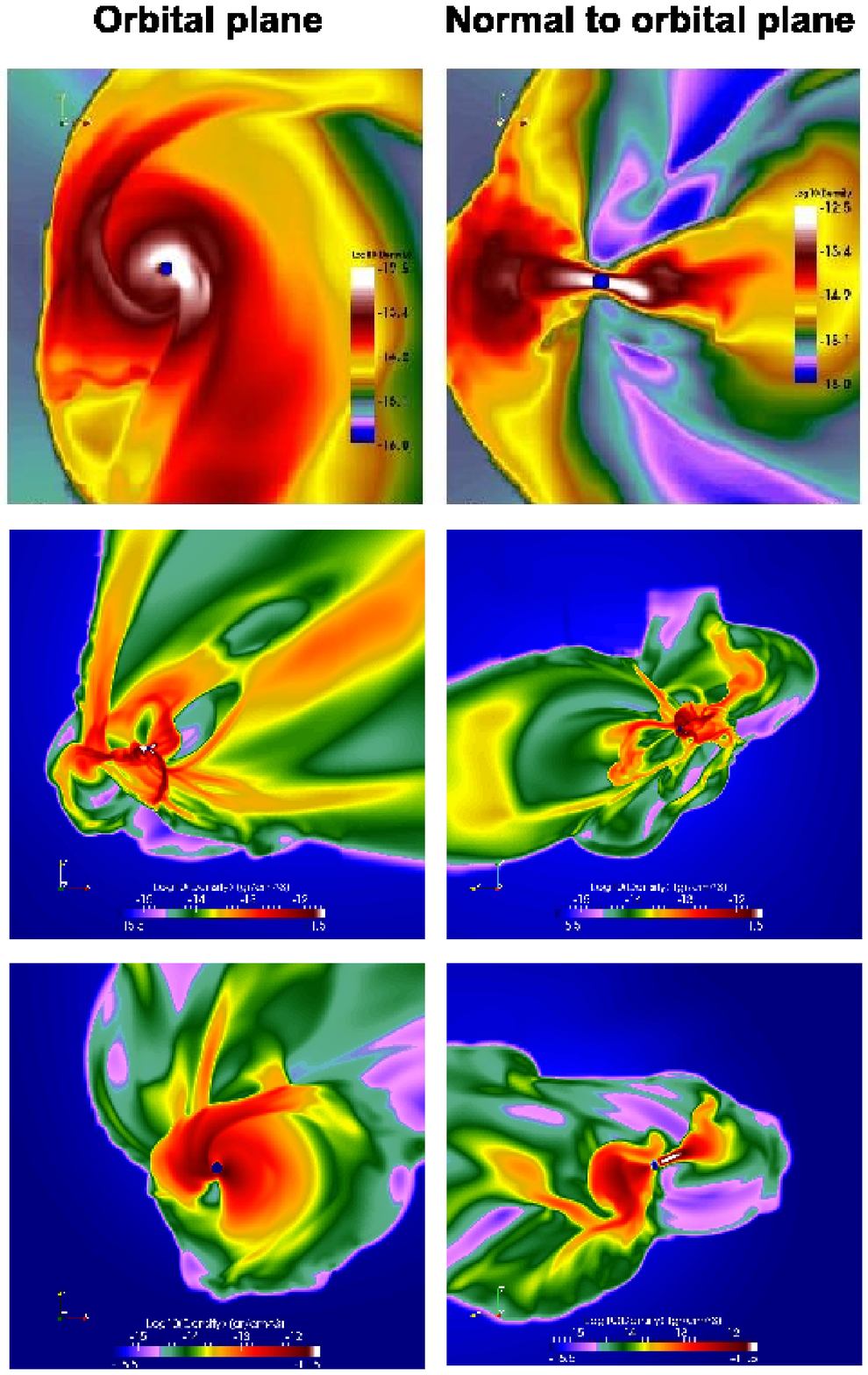}
}
\end{center}
\vspace{-0.8cm}
\caption{Accretion around the WD in the orbital plane (left column)
         and perpendicular to the orbital plane (right column) for
         $v_{\mathrm{RG}}=20$ (top row) and $v_{\mathrm{RG}}=60$
         (middle and bottom row). Shown is density, logarithmic
         scale. In the $v_{\mathrm{RG}}=20$ case, occurs permanently
         through a non-Keplerian disk with spiral shocks. In the
         $v_{\mathrm{RG}}=60$ case, accretion occurs predominantly
         through a 'turbulent accretion ball' (middle row) but
         occasionally also through a non-Keplerian disk (bottom row),
         similar to the $v_{\mathrm{RG}}=20$ case.  The spatial scale
         of each of the above panels is roughly $10^{13}$ cm.}
\label{fig:accretion}
\end{figure}

%
%
%
%
The fast and slow simulation also show substantial differences in the
accretion geometry around the WD, as shown in
Fig.~\ref{fig:accretion}. In the case of the slow RG wind, accretion
permanently takes place through a non-Keplerian disk in which angular
momentum is transported via spiral shocks. In the case of the fast RG
wind, the vicinity of the WD can rather be characterized as a
supersonically turbulent accretion ball. Matter flows in from various
directions and is redirected, slowed down, and thermalized by a
complicated and rapidly changing network of shocks, while angular
momentum is transported outwards. For a more thorough discussion of
supersonic turbulence in a non-periodic setting
see~\cite{2006A&A...459....1F}. The overall appearance of the vicinity
of the WD is much more fluctuating than in the slow case. Disk-like
accretion states also occur, but only as transient features.  In the
slow wind case, densities around the WD are up to two orders of
magnitude larger in the orbital plane than perpendicular to the
orbital plane~\citep{walder-folini-shore:08}. In the fast wind case no
such pronounced density contrast exists. The accretion rates change
from around 10\% $\dot{M}_{\mathrm{RG}}$ in the slow case to roughly
2\% in the fast case. By contrast, the accretion rate is a mere 0.7\%
in the adiabatic simulation we performed.

%
%
%
%
In all the cases considered here, the cumulative effect of the
system's mass and angular momentum losses points to a shrinking of the
binary orbit. The loss of angular momentum more than outweighs the
mass loss of the system and the two components spiral inwards.  This
picture is dominated by the quiescent phase of RS Oph, as only much
less matter (around 10\%) is involved in the occasional nova outburst.
Absolute numbers turn out, however, to depend strongly on the RG wind
velocity. Expressing the cumulated losses in terms of $da/a$ with $a$
the system separation, we find $da/a \approx 8 \cdot 10^{-8}$
yr$^{-1}$ in the slow case but about ten times less in the fast case,
$da/a \approx 8 \cdot 10^{-9}$ yr$^{-1}$. Of much less relevance here
are the thermodynamics. For $v_\mathrm{RG} = 20$~km/s and $\gamma =
5/3$ we find $da/a \approx 6
\cdot 10^{-8}$.
\section{Discussion and Conclusions}
\label{sec:discussion}
Our results further support the idea that the
recurrent nova RS Oph is a good candidate for a progenitor of an SN
Ia. In particular, the mass transfer rate
required by nova models to reach a net mass gain over several nova
cycles can be reached for reasonable system parameters. Also, the
orbit of RS Oph is found to shrink under current conditions. Chances
to maintain or even enhance the current mass transfer rate in the
future thus are intact.

However, the three critical time-scales which determine the ultimate
fate of the system are all of the same order. RG evolution models
predict a high mass loss for a RG for some million years only. This
time scale is similar to what is needed to shrink the orbit
significantly and to the time necessary for the WD to accrete enough
mass to become unstable.

To further investigate the possibility of RS Oph finally becoming a SN
Ia simulation results should cover the long term evolution of the
system and its components, over several nova cycles up to several
million years. In addition, multi-dimensional nova models with better
predictions of the mass gain condition of the WD would be most
welcome. On the observational side, better estimates on the ejected
mass in a nova outburst would be highly desirable to test the
consistency with existing nova models and related accretion physics.
\acknowledgements 
The authors wish to thank the crew running the HP Superdome at ETH
Zurich and the people of the Swiss Center for Scientific Computing,
CSCS Manno, where the simulations were performed.

%

\end{document}